\documentclass{article}

\usepackage{arxiv}

\usepackage[utf8]{inputenc} 
\usepackage[T1]{fontenc}    
\usepackage{hyperref}       
\usepackage{url}            
\usepackage{booktabs}       
\usepackage{amsthm}
\usepackage{amsmath}
\usepackage{amsfonts}       
\usepackage{nicefrac}       
\usepackage{microtype}      
\usepackage{cleveref}       
\usepackage{lipsum}         
\usepackage{graphicx}
\usepackage{natbib}
\usepackage{doi}

\usepackage[]{algorithm2e}
\usepackage{caption}
\usepackage{subcaption}

\graphicspath{{./figs/}}

\title{An adaptively optimized algorithm for counting nuclei in X-ray micro-CT scans of whole organisms}

\date{\today}

\author{ \href{https://orcid.org/0000-0002-4122-9800}{\includegraphics[scale=0.06]{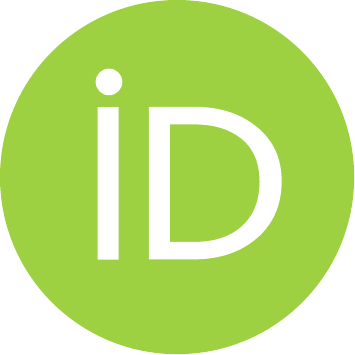}\hspace{1mm}Anna Madra}
\thanks{To whom correspondence should be addressed.} \\
	Department of Pathology\\
	The Pennsylvania State University\\
	Hershey, PA 17033 \\
	\texttt{annamadra@psu.edu} \\
	\And
	\href{https://orcid.org/0000-0002-1653-4168}{\includegraphics[scale=0.06]{orcid.pdf}\hspace{1mm}Alex YS.~Lin} \\
	Department of Pathology\\
	The Pennsylvania State University\\
	Hershey, PA 17033 \\
	\texttt{ayl114@psu.edu} \\
	\And
	\href{https://orcid.org/0000-0002-9221-8634}{\includegraphics[scale=0.06]{orcid.pdf}\hspace{1mm}Daniel J.~Vanselow} \\
	Department of Pathology\\
	The Pennsylvania State University\\
	Hershey, PA 17033 \\
	\texttt{dxv46@psu.edu} \\
	\And
	\href{https://orcid.org/0000-0002-5350-5825}{\includegraphics[scale=0.06]{orcid.pdf}\hspace{1mm}Keith C.~Cheng} \\
	Department of Pathology\\
	The Pennsylvania State University\\
	Hershey, PA 17033 \\
	\texttt{kcheng76@gmail.com}
}

\renewcommand{\shorttitle}{\textit{arXiv} Template}

\hypersetup{
pdftitle={Adaptive algorithm for counting nuclei in organisms},
pdfsubject={Original research article},
pdfauthor={Anna Madra, Alex YS.~Lin, Daniel J.~Vanselow, Keith C.~Cheng},
pdfkeywords={X-ray microCT, X-ray histotomography, quantitative analysis, reduced-order modeling, image processing},
}

\begin{document}
\maketitle

\begin{abstract}
Living organisms are primarily made of cells. Identifying them and characterizing their geometry and spatial distribution is a first step towards building multi-scale models of these biomaterials. We propose a method to count cells using nuclei in an X-ray microtomographic scan of a zebrafish. To account for scanning artifacts and partial volume effect, the method is adaptively calibrated using parameters approximated from the manifold of manually selected and optimized special cases. The methodology is tested on nuclei in the eyes of zebrafish larvae of different ages.
\end{abstract}

\keywords{X-ray microCT \and X-ray histotomography \and Quantitative analysis \and Reduced-order modeling \and Image processing}

\section{Introduction}
Compared to engineering materials, biomaterials and living organisms are highly heterogeneous and form stochastically varying hierarchical structures at multiple length scales. Fortunately, as postulated by Matthias Schleiden for plants \citep{Schleiden1838} and Theodor Schwann for animals \citep{Schwann1839}, they all have a unit building block: a cell. Hence, to anchor the molecular data and understand the changes in development and mechanical behavior of biological materials, it is essential to identify cells, characterize their geometry, and observe their spatial distribution. Particularly, spatial distribution helps to determine the formation of organs -- natural, highly heterogeneous composites.\newline

In \citep{Ding2019,Katz2021}, a modified setup of X-ray microtomography combined with specimen preparation protocols, together referred to as X-ray histotomography, has been used to image the entire organism at cellular resolution. The X-ray histotomography is unprecedented in the field of biology that mainly relies on 2.5D histology slides to provide information on the morphology of biomaterials. It also opened the possibility to create a first quantitative cellular model of an organism. The initial step towards creating this model consists of cell identification and counting.\newline

Several methods can be used to identify cells. Most rely on the identification of the border of the cell, either by using gradient maps \citep{Stringer2021}, watershed transform \citep{Lux2020}, or by altering specimen preparation protocol to enhance cell membranes \citep{Dimopoulos2014} and thus simplify image processing. Recently, with the advances in deep learning, many approaches have been proposed that rely on Convolutional Neural Networks (CNN) and their derivatives \citep{Vu2019,Lux2020,Scherr2020,Stringer2021}. While these methods continue to improve in the accuracy of detection, even when applied to general cases, like in \citep{Stringer2021}, they still require huge, manually annotated datasets to train or update the model. Such databases are becoming publicly available \citep{Rosenhain2013,Edlund2021}, albeit their focus is on high-resolution 2D data from optical end electron microscopy.\newline

Overall, the existing approaches have two drawbacks
\begin{enumerate}
\item They rely on high-resolution data where pixel/voxel size is at least one tenth of the cell size, although one-hundredth is more common.
\item They are optimized for two-dimensional data sets.
\end{enumerate}
The first drawback is critical from the computational point of view, since the highest resolution for the large field-of-view X-ray microCT imaging still leads to nuclei diameters equal to 5-6 voxels. Hence, most automated segmentation methods like thresholding or methods involving gradient maps will not be efficient. Methods involving the watershed algorithm may be effective but are still prone to errors due to scan artifacts, partial volume effect in particular. Thus, they require local optimization to improve performance.\newline

The second drawback -- focus of the most existing methods on 2D data sets, is of lesser importance computationally, but essential in biology. Histology -- light microscopy using tissue slabs, which is fundamentally 2D, has been the de facto central reference point for biologists who use aspects of tissue structure as scientific output for over 100 years. The 2.5D histology imaging requires removal of material in adjacent slices thus some of the information on cell distribution is lost. Fortunately, most of the image processing methods can be generalized into 3D. The challenge here is the difficulty of validation. While comparison of the results in 2D is relatively straightforward for both a non-expert and a pathologist or a biologist, it is not the case for 3D data. Even establishing a "ground truth" manually segmented data set is prone to variation depending on the operator's skill, usually requiring substantial workload and specialized equipment, like a dedicated Virtual Reality hardware and software \citep{Syglass}. Given these operational difficulties, it is crucial to limit the number of cases for which the algorithm has to be manually calibrated.\newline

In our method, we address both of these drawbacks by proposing an optimization algorithm to count the cells using nuclei as a proxy, but calibrating it using a manifold of parameters obtained based on expert-validated special cases. Here we describe a nucleus counting algorithm, followed by the adaptive parameterization, and the results for counting nuclei in the left and right eyes of zebrafish larvae (\textit{Danio rerio}) at two different stages of development. In conclusion, we hoped to provide a perspective on how this approach can be expanded to facilitate further characterization of other specimens of biological origin.

\section{Nucleus counting algorithm}
\begin{figure}
  \centering
  \includegraphics[width=0.8\textwidth]{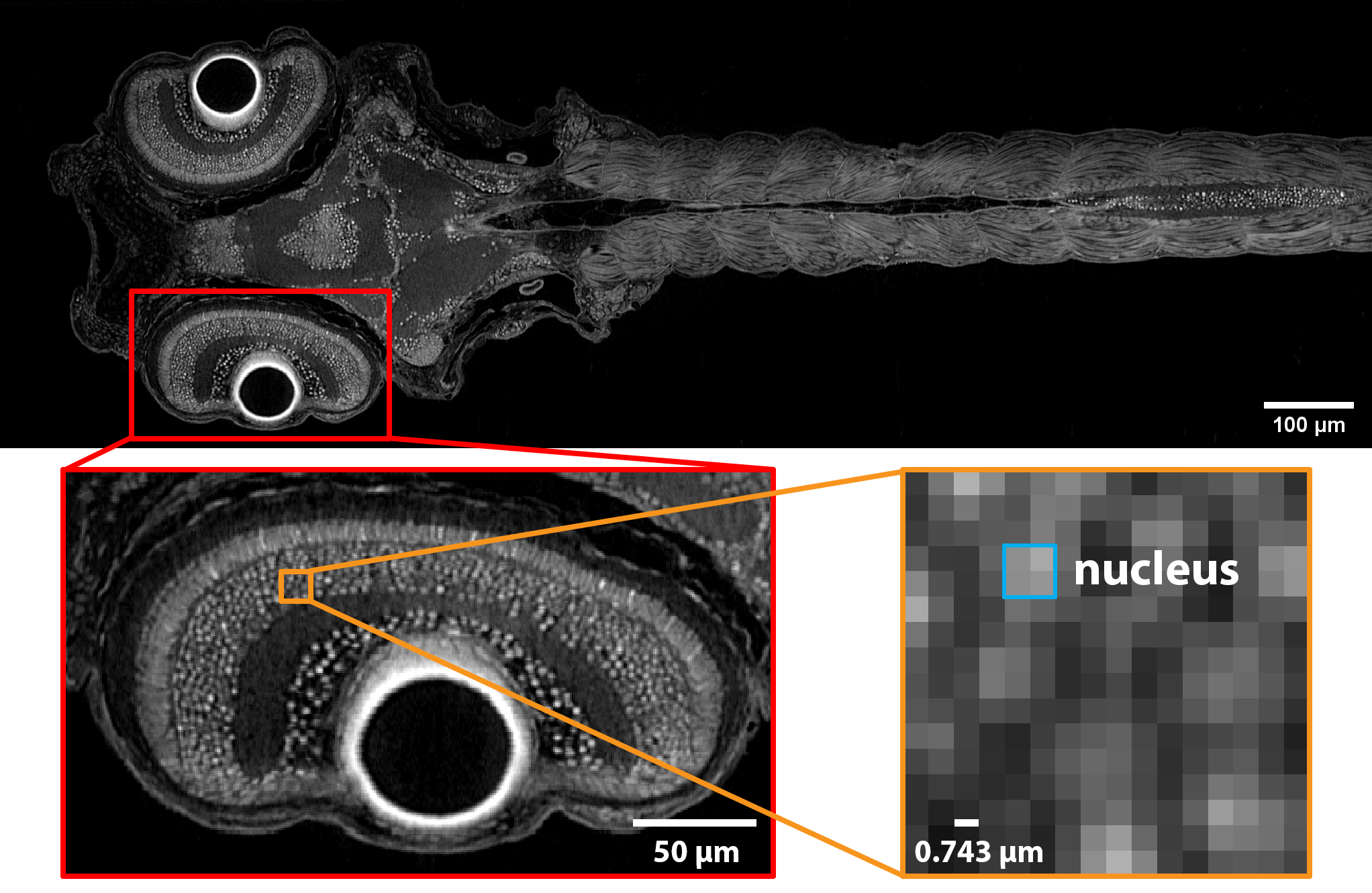}
  \caption{A 2D slice of X-ray histotomgraphy of a 5 dpf (days post fertilization) zebrafish (\textit{Danio rerio}). Insets show the left eye and the nuclei of interest compared to voxel size.}
  \label{fig:zebrafish}
\end{figure}

We chose to concentrate on the cells in the eye of the zebrafish (Fig. \ref{fig:zebrafish}) since, apart from the incomplete mitosis, these cells contain only one nucleus that is not in physical contact with another \citep{Hildebrand2017}. Thus, identification of a nucleus is equivalent to a detection of a single cell. In the X-ray histotomographic scans, the nuclei are represented by groups of voxels with locally higher intensities, indicating higher X-ray attenuation. These groups are similar in shape to slightly irregular, almost spherical ellipsoids. At the resolution in question: 0.743 $\mu$m, they span 5-6 voxels in perpendicular directions. Since the goal of this study is only to estimate the number of nuclei, and through them, cells, and not to segment the data set, the problem can be reduced to finding extrema in the 3D voxel intensity function within a neighborhood of size $X\times Y\times Z$ with a minimum distance $D$ between results.\newline

The solution to this problem is described in Alg. \ref{alg:maxima}. First, we inspect the 3D subsets of size $X\times Y\times Z$ and return the maxima within each. Then, we remove the maxima that fall below the threshold $T$ of minimum voxel intensity, and finally, the maxima that are within $D$ distance of each other are averaged into one result.

\begin{algorithm}[H]
\label{alg:maxima}
 \KwData{3D array of voxel intensities, sizes $X$, $Y$, $Z$, distance $D$, threshold $T$}
 \KwResult{number of nuclei $N$}
 \vspace{0.35cm}
 \For{all voxel groups of size $X\times Y\times Z$}{
  find coordinates of voxel with max intensity\;
  }
  \vspace{0.35cm}
 \For{all found maxima}{
 \uIf{maximum > $T$ and no maximum exists within $D$}
  {add 1 to $N$\;}
 \uElseIf{maximum > $T$}
  {average maximum with neighbors\;
  remove neighbors from maxima list\;
  add 1 to $N$\;}
 \uElse{
  remove from maxima list\;
 }
 }
 \caption{Nucleus counting.}
\end{algorithm}

\section{Adaptive calibration with the manifold of parameters}

\begin{figure}
  \centering
  \includegraphics[width=0.8\textwidth]{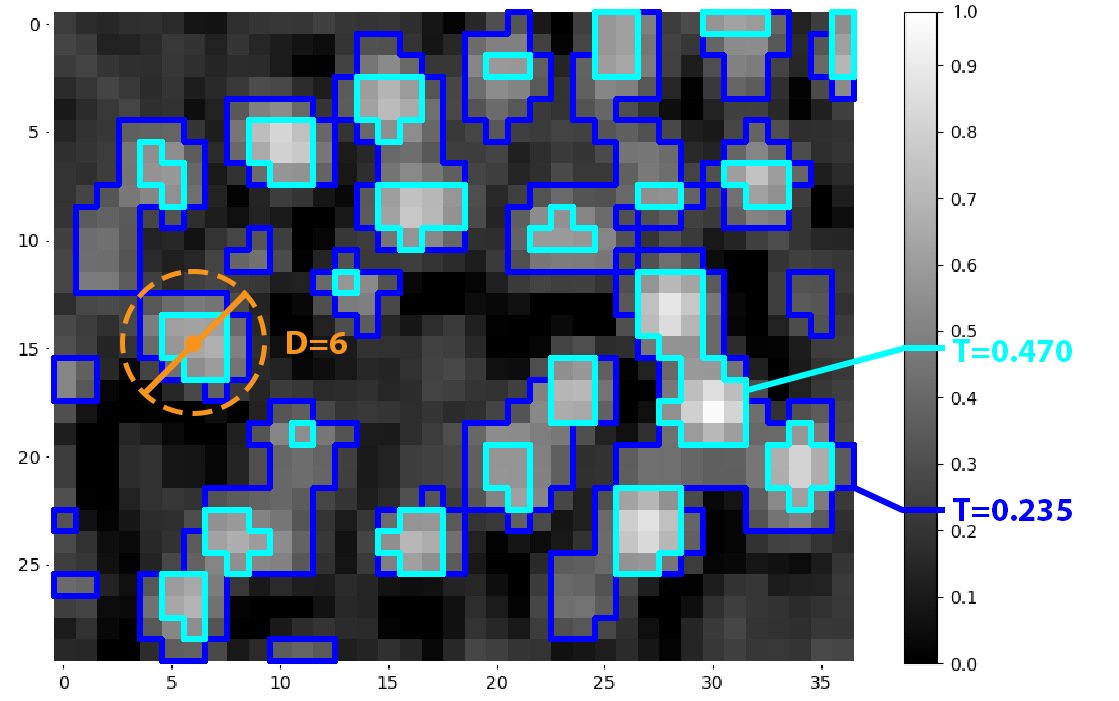}
  \caption{Parameters of the algorithm: intensity threshold $T$ and distance $D$.}
  \label{fig:threshold}
\end{figure}

The nucleus counting algorithm described in the previous section is straightforward, but its performance heavily relies on the choice of parameters $X$, $Y$, $Z$, $D$, and $T$. X-ray microCT data is subject to partial volume effect when the size of the object of interest is approaching the scan resolution. This means that objects imaged at a higher resolution consist of areas with homogeneous voxel intensity and have distinct borders, while at lower resolutions, the homogeneous area of voxel intensity is surrounded by a fringe of the average of the object's intensity and its surroundings. This implies that if the image in question contains only nuclei and homogeneous background, then the choice of intensity threshold $T$ is equivalent to the designation of the border of a nuclei area (Fig. \ref{fig:threshold}).

Thus we can reduce the parameters considered to $T$ and $D$. While this is a manageable number to optimize for computations, it is nevertheless open to interpretation. For example, for the left eye in a 5 dpf fish specimen, manual adjustment resulted in a range of nuclei counts from 50 to 72 thousand nuclei, implying an almost 30 percent difference. While the choice of $D$ is intuitive and can be easily computed given the maximum sizes of nuclei, the choice of $T$ depends on the following:

\begin{enumerate}
\item Intensity values in the scan, depending on the X-ray microCT setup, specimen geometry, and reconstruction algorithm.
\item Resolution of the scan and the prevalence of the partial-volume effect.
\item Spacing of the nuclei.
\end{enumerate}

While the first two can be controlled and standardized, it is not yet practicable for exploratory X-ray microCT scans. The third varies throughout the specimen (Fig. \ref{fig:threshold}) and in areas of high nuclei density can cause identification of false positives. In those areas, the intensity threshold $T$ needs to be higher to prevent the merging of neighboring nuclei into one central "nucleus".

\begin{figure}
  \centering
  \begin{subfigure}[b]{0.33\textwidth}
  \centering
  \includegraphics[width=\textwidth]{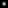}
  \caption{individual nucleus}
  \label{fig:case1}
  \end{subfigure}
  \begin{subfigure}[b]{0.33\textwidth}
  \centering
  \includegraphics[width=\textwidth]{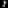}
  \caption{two nuclei overlapping in the same plane}
  \label{fig:case2}
  \end{subfigure}
  \begin{subfigure}[t]{0.33\textwidth}
  \centering
  \includegraphics[width=\textwidth]{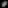}
  \caption{two nuclei overlapping axially}
  \label{fig:case3}
  \end{subfigure}
  \begin{subfigure}[t]{0.33\textwidth}
  \centering
  \includegraphics[width=\textwidth]{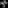}
  \caption{four nuclei overlapping in multiple planes}
  \label{fig:case4}
  \end{subfigure}
  \caption{Special cases: a) individual nucleus; b) two nuclei overlapping in the same plane; c) two nuclei overlapping axially; d) four nuclei overlapping in multiple planes. NB: here, overlap does not indicate connection -- the nuclei are always physically separated. The overlap is due the partial-volume effect that turns the image into slab projections.}
  \label{fig:cases}
\end{figure}

To consider these special cases, we compiled a data set of representative voxel regions. Here, to ease comprehension, we show examples of their 2D equivalents in Fig. \ref{fig:cases}. The optimal values of $T$ were then determined by a human expert for each special case.

The key part of the algorithm is determining which case is represented by the data and then looking up the $T$ value. To measure the similarity of the analyzed 3D subset of the data and special cases, we propose to compare principal components of their snapshots $s_m$
\begin{equation}
s_m=\begin{bmatrix}
\vdots\\
\mu (x,y,z)\\
\vdots
\end{bmatrix}\in \mathbb{R}^N
\end{equation}
where $\mu$ is the X-ray intensity of a voxel at a location $x$, $y$, $z$ and $m=1,\ldots,M$ with $M$ -- number of expert-optimized special cases. The snapshots are collected in a matrix $S$
\begin{equation}
S=\begin{bmatrix}
\\
s_1&\cdots&s_m&\cdots&s_M\\
&
\end{bmatrix}\in\mathbb{R}^{N\times M}
\end{equation}
The principal components are then calculated by performing a Singular Value Decomposition (SVD) of the snapshot matrix
\begin{equation}\label{eq:svd}
S=U\Sigma V^*
\end{equation}
where $\Sigma$ is a diagonal matrix of singular values and $U$, $V$ are orthonormal bases of snapshots and
\begin{equation}
U=\begin{bmatrix}
\\
u_1&\cdots&u_M\\
&&
\end{bmatrix}.
\end{equation}

\begin{figure}
  \centering
  \includegraphics[width=\textwidth]{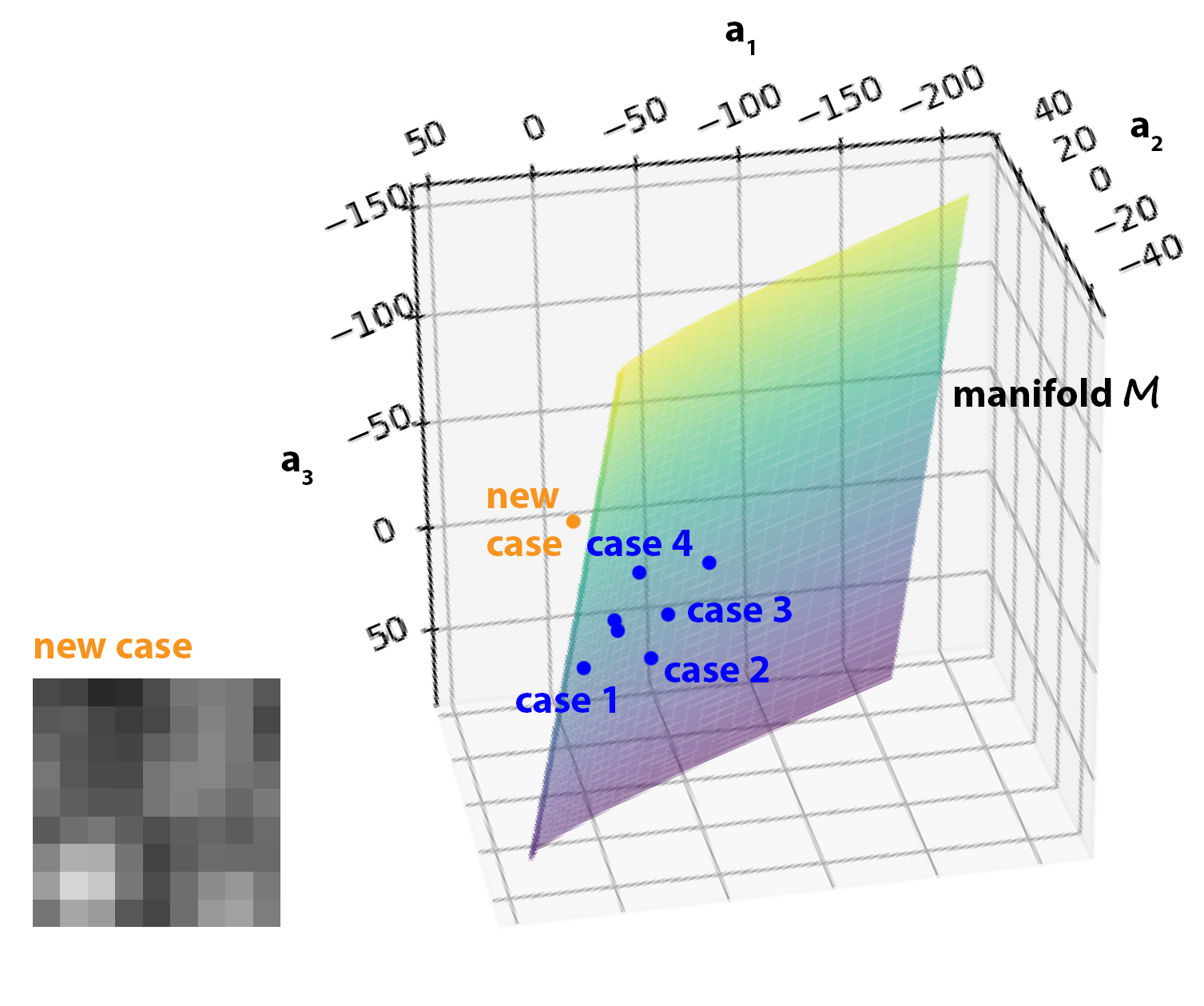}
  \caption{Manifold of the special cases with the arbitrary new case projected into the reduced-order feature space. The manifold interpolates parameters for intermediary cases based on the calibration performed by the biologist.}
  \label{fig:manifoldRecon}
\end{figure}

The optimized special cases are projected to the new base
\begin{equation}
A=\begin{bmatrix}
\\
\mathbf{a}^{(1)}&\cdots&\mathbf{a}^{(R)}\\
&
\end{bmatrix},\ \mathbf{a}^{(r)}=\begin{bmatrix}
a_1^{(r)}\\
\vdots\\
a_R^{(r)}
\end{bmatrix}
\end{equation}

with the first two principal components shown in Fig. \ref{fig:manifoldRecon} along with the associated optimal algorithm parameters. It can be observed that parameter distribution is non-linear, hence we propose to approximate it by assuming a low-dimensional manifold $\mathcal{M}$
\begin{equation}
\mathcal{M}(\mathbf{a})=0,\ \mathcal{M}\in\mathbb{R}^R,
\end{equation}
approximated as
\begin{equation}
\psi(\mathbf{a})=p^T(\mathbf{a})c(\mathbf{a})
\end{equation}
where $p$ are polynomial basis functions and $c$ are coefficients minimizing the weighted moving least squares criterion \citep{Nayroles1992,Breitkopf2000,Raghavan2013,Madra2018}
\begin{equation}
J(\mathbf{a}(a))=\frac{1}{2}\sum_{\mathbf{a}^{(r)}\in V(\mathbf{a})}w(\mathbf{a}^{(r)},\mathbf{a})(p^T(\mathbf{a}^{(r)})c-\mathbf{a}^{(r)})^2
\end{equation}
and $V(\mathbf{a})$ is a neighborhood defined by the weighing function
\begin{equation}
w(\mathbf{a}^{(r)},\mathbf{a})=\mathrm{exp}\left(-\frac{||\mathbf{a}^{(r)}-\mathbf{a}||}{2d^{(r)}}\right)
\end{equation}
where $d^{(r)}$ is the radius of influence of point $\mathbf{a}^{(r)}$.\newline

To determine optimal parameters of the counting algorithm, we take the voxel group $X\times Y\times Z$, project it to the new base obtained in Eq. \ref{eq:svd}, and find the nearest point on the manifold of parameters (Fig. \ref{fig:manifoldRecon})
\begin{equation}
\mathbf{a}^*=\mathrm{Argmin}(\mathrm{dist}(\mathcal{M},\mathbf{a}^{(r)})).
\end{equation}
The parameters at this point are then used to calibrate the counting algorithm. To further increase the accuracy of the approach, the manifold can be refined by adding new expert-validated cases, guided by the geometry of cases back-projected from the manifold.

\section{Results}
The nucleus counting algorithm has been applied to four datasets: left and right eye of a 3 dpf (days post fertilization), and the left and right eye of a 5 dpf zebrafish larvae (\textit{Danio rerio}) scanned at a 0.743$\mu m$ resolution. Detailed description of the specimen acquisition, scanning setup, and reconstruction is described in \citep{Ding2019}.\newline

\begin{table}[htp]
\centering
\caption{Results of the nuclei count before and after adaptive optimization.}
\label{tab:res}
\begin{tabular}{@{}lccc@{}}
\toprule
Specimen        & \begin{tabular}[c]{@{}c@{}}Nuclei count with\\ default parameters\end{tabular} & \begin{tabular}[c]{@{}c@{}}Nuclei count after\\ optimization\end{tabular} & Change \\ \midrule
left eye, 3dpf  & 26 765 & 30 095 & +12.4\% \\
right eye, 3dpf & 24 294 & 29 520 & +21.5\% \\
left eye, 5dpf  & 51 106 & 68 297 & +33.6\% \\
right eye, 5dpf & 54 334 & 63 765 & +17.4\% \\ \bottomrule
\end{tabular}
\end{table}

\begin{figure}
  \centering
  \begin{subfigure}[b]{0.54\textwidth}
  \centering
  \includegraphics[width=\textwidth]{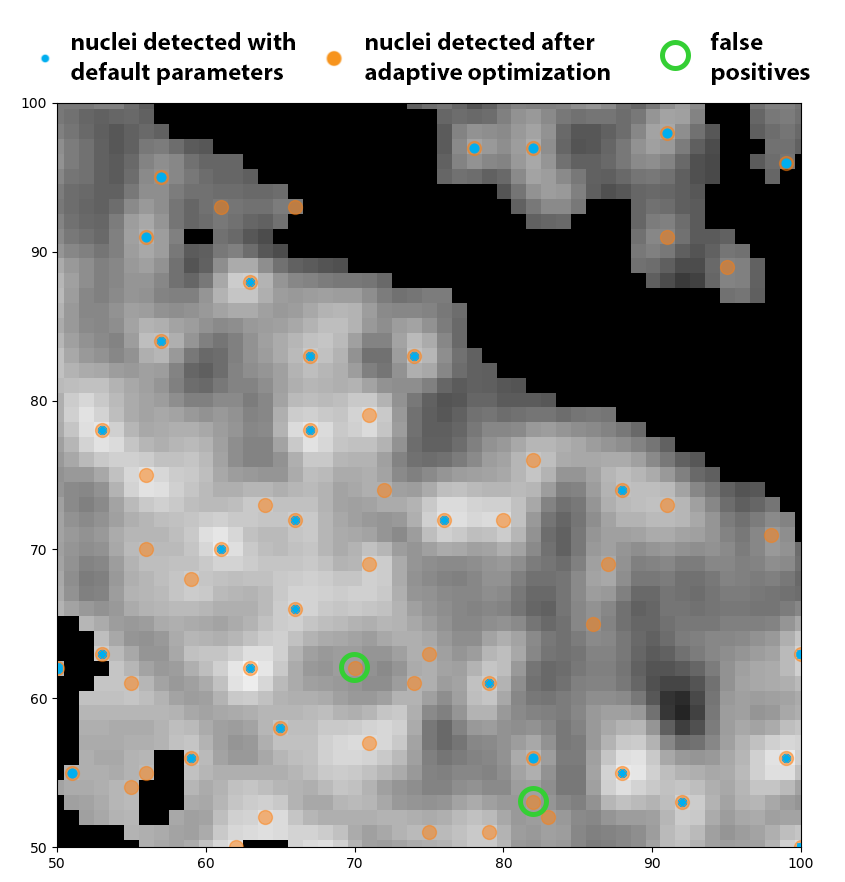}
  \caption{2D results before and after optimization}
  \label{fig:2Dres}
  \end{subfigure}
  \begin{subfigure}[b]{0.45\textwidth}
  \centering
  \includegraphics[width=\textwidth]{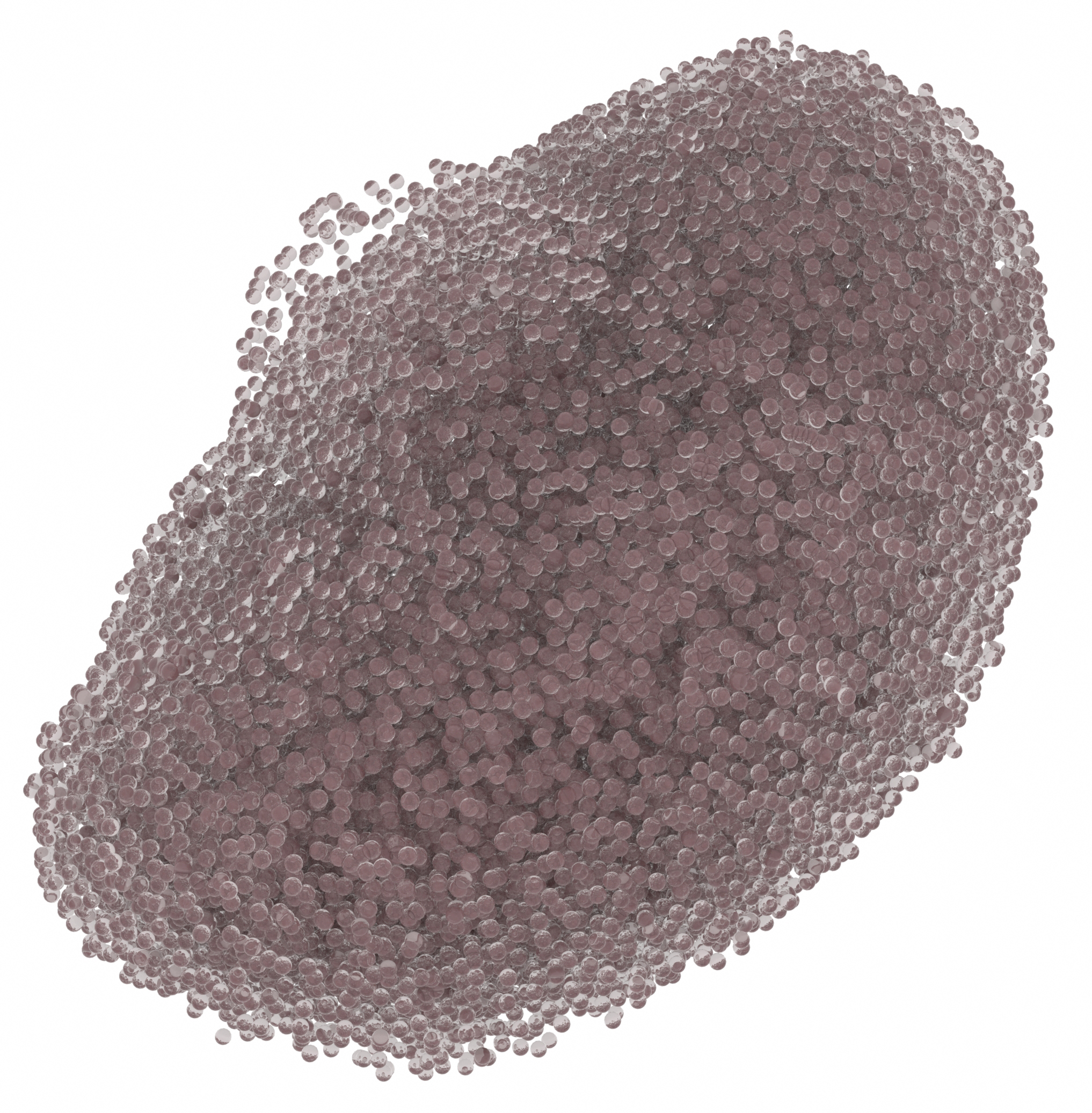}
  \caption{3D rendering of all detected nuclei}
  \label{fig:3D}
  \end{subfigure}
  \caption{a) Comparison of the 2D results for nonoptimized and adaptively optimized nucleus counting algorithm. b) 3D rendering of nuclei detected in the left eye of the 5 dpf zebrafish.}
  \label{fig:results}
\end{figure}

The nuclei count before and after calibration is summarized in Table~\ref{tab:res}. Without adaptive calibration, the nuclei count was lower by at least ten percent, reaching over thirty. The default parameters are selected manually, based on a visual appraisal of the accuracy of the count, using samples like the one shown in Fig. \ref{fig:2Dres} for 2D data. But even for a small 2D sample, the heterogeneity of the data set caused 28 nuclei to remain undetected, with two false positives identified post optimization. After optimization with the initial set of manually-calibrated eight special cases the count was more accurate, although a thorough validation and further refinement of the method are pending. The nuclei locations detected so far have been reconstructed in 3D reconstruction shown in Fig. \ref{fig:3D}. These 3D renders will further assist content experts in tissue structure in validating the method from the biological standpoint.

\section{Conclusions}
The proposed approach to calibrate a simple nucleus counting algorithm has two advantages over more traditional approaches. On the one hand, it manages the inconsistency and subjectivity introduced when counting is done by a human expert. On the other, it also limits the number of special cases that need to be considered. While it can be argued that this problem can be successfully solved by training a Deep Learning algorithm, our approach empowers non-computational domain experts, allowing them to validate special cases and further refine the solution as necessary. This control over the performance of the algorithm is crucial. It ensures that the quality of the results can be assessed and improved through case validation, a process much more common in medicine and biology, rather than algorithm parameterization and training, ubiquitous in machine learning. We expect that an interdisciplinary approach combining expertise in the materials being imaged with more accessible control of the algorithmic processing will result in a higher accuracy of nuclei counts.\newline

Currently, this methodology for nucleus counting is being thoroughly validated from the biological standpoint and tested on additional cell types in zebrafish specimens. The approach of an algorithm calibrated by a manifold of parameters is universal and could find applications in other fields, from locally adapted X-ray microCT segmentation to local refinement of FEM meshes, everywhere the input of human expert is required to "fix" the algorithm.

\bibliographystyle{unsrtnat}
\bibliography{manifoldCounting2}
\end{document}